\documentclass{article}
\usepackage{graphicx} 
\usepackage{amsmath,amsfonts,amssymb, amsthm}
\usepackage{setspace}
\usepackage{tocloft}
\usepackage{hyperref}
\usepackage{float}
\usepackage{authblk}
\usepackage{algorithm}
\usepackage{algpseudocode}
\usepackage{float}
\usepackage{pgfplots}
\pgfplotsset{compat=1.18}
\usepackage{pgfplotstable}

\newtheorem{definition}{Definition}
\newtheorem{theorem}{Theorem}

\providecommand{\keywords}[1]
{
  \small	
  \textbf{Keywords: } #1
}

\title{Post-Quantum Stealth Address Protocols}
\author[1]{Marija Mikić}
\author[2]{Mihajlo Srbakoski}
\author[3]{Strahinja Praška}
\affil[1,2 ]{Faculty of Mathematics, University of Belgrade}
\affil[3]{Faculty of Technical Sciences, University of Novi Sad}
\date{January 2025}

\begin{document}

\maketitle

\begin{abstract}
The Stealth Address Protocol (SAP) allows users to receive assets thro\-ugh stealth addresses that are unlinkable to their stealth meta-addresses. The most widely used SAP, Dual-Key SAP (DKSAP), and the most performant SAP, Elliptic Curve Pairing Dual-Key SAP (ECPDKSAP), are based on elliptic curve cryptography, which is vulnerable to quantum attacks. These protocols depend on the elliptic curve discrete logarithm problem, which could be efficiently solved on a sufficiently powerful quantum computer using the Shor algorithm. In this paper three novel post-quantum SAPs based on lattice-based cryptography are presented: LWE SAP, Ring-LWE SAP and Module-LWE SAP. These protocols leverage Learning With Errors (LWE) problem to ensure quantum-resistant privacy. Among them, Module-LWE SAP, which is based on the Kyber key encapsulation mechanism, achieves the best performance and outperforms ECPDKSAP by approximately 66.8\% in the scan time of the ephemeral public key registry.
\end{abstract}

\keywords{Blockchain, privacy, stealth address, lattice-based cryptography, learning with errors, Kyber}

\section{Introduction}

The protection of user privacy in blockchain transactions is of paramount importance. Stealth address protocols (SAP) allow assets to be received via stealth addresses that are not directly linked to the corresponding stealth meta-addresses. Various cryptographic methods can be used to generate SAP. The DKSAP protocol \cite{textbook1} utilizes elliptic curve multiplication in combination with a hashing of the resulting shared secret, while ECPDKSAP \cite{MM} uses an alternative method involving elliptic curve pairing. In addition, SAP can also be derived using Post-Quantum Cryptography (PQC) approaches such as lattice-based cryptography, Learning With Errors (LWE), Ring Learning With Errors (Ring-LWE) and Module Learning With Errors (Module-LWE). Although blockchain technology has not yet integrated PQC, these protocols are expected to play a crucial role in improving the privacy of blockchain transactions in the near future.

Stealth address is a method of obtaining addresses using key management with privacy. A recipient generates the private key and publishes the public key, which is called the stealth meta-address. Each transaction (stealth) address for this user can be obtained from this stealth meta-address. The corresponding stealth address private key can be easily calculated from the corresponding private key of the stealth meta-address and a public ephemeral key (only the recipient can calculate the stealth address private key). It is imperative for the recipient to maintain the privacy of the stealth meta-address's private key, as it is essential for the recipient to identify and access their transactions recorded in the blockchain ledger.

The problem with DKSAP and ECPDKSAPs is that their security relies on the elliptic curve discrete logarithm problem. This problem could easily be solved on a sufficiently powerful quantum computer running the Shor algorithm \cite{shor} or even faster alternatives.

The basic algebraic structure used in post-quantum cryptography is the lattice. The Shortest Vector Problem (SVP) is the most important computational problem in lattice-based cryptography. It presents us with the task of determining the minimum Euclidean length of a non-zero vector of the lattice. The LWE problem has a direct connection to SVP. This connection emerges from the fact that solving the LWE problem is at least as hard as certain worst-case lattice problems, including SVP. Variants such as Ring-LWE (RLWE) and Module-LWE (MLWE) extend the LWE problem to structured lattices, leveraging their algebraic properties to achieve more efficient cryptographic constructions. In recent years, lattice-based cryptography has been recognized for its many attractive properties, such as strong provable security guarantees and apparent resistance to quantum attacks, flexibility in realizing powerful tools such as fully homomorphic encryption, and high asymptotic efficiency. Indeed, several works have shown that for basic tasks such as encryption and authentication, lattice-based primitives can have performance competitive with (or even outperform) those based on classical mechanisms such as RSA or Elliptic Curve Cryptography (ECC). On the other hand, lattice-based schemes have much larger keys and ciphertexts in comparison, which can be a disadvantage in resource-constrained environments. In lattice-based cryptography, for example, the keys are often several kilobytes in size, whereas ECC keys are usually only 32 bytes long.

In this paper, we introduce three LWE protocols: LWE Stealth Address Protocol, RLWE Stealth Address Protocol and MLWE Stealth Address Protocol. We compare the parsing time of these protocols. It is important to mention that our two LWE Stealth Address Protocols (RLWE and MLWE SAPs) provide more powerful results in terms of address computation time compared to the results of DKSAP \cite{textbook1} and ECPDKSAP - Protocol 3 from \cite{MM}.

The paper consists of seven sections. The first section is the introduction. The second section provides an overview of the main literature on stealth addresses. $\mbox{Section 3}$ introduces us to the post-quantum hard problems. This section is divided into four subsections. In the first subsection, we introduce the notation that we will use in the rest of the paper. We also define the concept of the lattice, the ideal lattice, the SVP and the Short Integer Solution (SIS). In the remaining subsections of this section, we focus on the LWE problems. We define both the Search-LWE and Decision-LWE problems as well as their variants: Search-RLWE, Decision-RLWE, Search-MLWE and Decision-MLWE. In the fourth section, we describe the structure inherent in each Stealth Address Protocol (SAP). We also describe concepts such as the view tag (subsection 4.1) and the viewing key (subsection 4.2), which are standard in the DKSAP and ECPDKSAPs protocols mentioned earlier and in the protocols presented in this paper. Section 5 is devoted to Learnig With Errors Stealth Address Protocols. This section is the centerpiece of our research. In this section, we have five subsections. In the first subsection, we introduce the notation that we use in the rest of the paper, as well as the basic functions. In $\mbox{subsection 2}$, we introduce Kyber, the post-quantum safe key encapsulation mechanism (KEM) that we use for sharing secrets in our MLWE protocol. Subsection 5.3 provides a detailed description of our Module Learning With Errors Stealth Address Protocol (MLWE SAP). In the following subsections, we describe our Ring Learnig With Errors Stealth Address Protocol (RLWE SAP) and Learnig With Errors Stealth Address Protocol (LWE SAP). The reason why we have chosen this arrangement of sections (and not the other way around) is that the MLWE SAP is the most efficient protocol compared to these two protocols and we want to focus on it the most. In Section 6, we have three subsections. In $\mbox{subsection 6.1}$ we compare the efficiency of the PQ SAPs (LWE SAP, RLWE SAP and MLWE SAP). In this part of the paper, we also compare the time required for the recipient to compute the address for MLWE SAP without the view tag and with the view tag (considering both a 1-byte view tag and a full hash derived from the shared secret). In subsection 6.2, we compare MLWE SAP (most efficient LWE SAP) with the most efficient ECPDKSAP. In the last subsection, we compare different parameter values for different KEMs. Section 7 is the conclusion.

\section{Related work}

The history of stealth addresses in cryptocurrencies began around the year 2013 with conceptual proposals aimed at improving the privacy of transactions. In 2013, Nicolas van Saberhagen described the Crypto\-No\-te protocol \cite{textbook20}, which used stealth addresses to improve the privacy of blockchain transactions. In 2014 Peter Todd \cite{textbook19} proposed integrating stealth addresses into the Bitcoin ecosystem to protect the privacy of receiving funds. The first prominent implementation of stealth addresses was in Monero \cite{textbook12}, which was launched in April 2014. Monero was designed from the ground up with privacy as a core feature, employing advanced cryptographic techniques to obscure transaction details and protect user anonymity. Stealth addresses play a crucial role in protecting the recipient's privacy by generating one-time use addresses for every transaction. To protect the identity of the sender, Monero utilizes ring signatures \cite{ringsign}, a technique in which the sender's transaction is mixed with multiple decoy transactions. When a user initiates a transaction, their signature is included in a "ring" of signatures, making it computationally infeasible for an observer to determine which participant is the actual sender. Monero also uses Ring Confidential Transactions (RingCT) to hide the transaction amounts. By using range proofs, RingCT \cite{ringct} ensures that the transaction values are within valid limits without revealing the actual amounts transferred.

The Dual-Key Stealth Address Protocol (DKSAP) \cite{textbook1,textbook4}, introduced in 2014, is the first stealth address protocol that, in addition to the spending key, also contains a viewing key, which is necessary for regulatory bodies to inspect transactions. The viewing key is a private key that is used to access all stealth addresses of a user, but without the possibility to spend funds from those addresses. With DKSAP, the recipient must search the ephemeral public key registry until a matching ephemeral public key is found. In DKSAP, after extracting the ephemeral public key, the recipient usually needs to perform two elliptic curve multiplications, two hash operations and one elliptic curve point addition to calculate stealth address. One technique that allows recipients of stealth address transactions to skip certain steps in the parsing process, and thus speed it up, is to add a view tag to each ephemeral public key. The view tag implemented in Monero \cite{textbook12}, on average, enables all steps in the parsing process to be performed only in $\frac{1}{256}$ cases, that is, only if the view tag matches. In other cases, it is necessary to calculate only the view tag, which usually requires only 1 hash operation and 1 elliptic curve multiplication. In this way, in \cite{textbook1}, parsing is accelerated by almost 87\%. 

The Zerocash protocol \cite{textbook7}, introduced in 2014, enables private transactions through a shielded pool where users deposit funds. It introduces "notes," cryptographic tokens representing value, each embedding a hidden amount, an owner key, and a nullifier to prevent double spending. Using ZK-SNARKs (zero-knowledge succinct non-interactive arguments of knowledge), Zerocash validates ownership and enforces transaction rules without exposing sensitive details, ensuring secure and private value transfer. Zerocash serves as the foundation of Zcash \cite{zcash}, a cryptocurrency that incorporates its privacy features directly into its blockchain. Unlike Tornado Cash \cite{tornado}, which acts as a mixer on existing blockchains by obfuscating links between deposits and withdrawals, Zerocash achieves systemic anonymity via shielded transactions embedded in its core protocol. Tornado Cash has been banned, and its founders were arrested due to allegations of facilitating money laundering and failing to implement mechanisms to prevent illegal usage. In contrast, Zcash enables "selective disclosure," allowing users to share transaction details when needed for purposes such as regulatory audits or tax compliance. 

The paper \cite{textbook1} explores the evolution of stealth addresses, addressing challenges such as DoS attacks and de-anonymization while proposing solutions to mitigate these issues. It introduces BaseSAP as a foundational framework for developing various stealth address schemes, including pairing-based and lattice-based stealth address protocols. The SAP protocols that use pairing are discussed in \cite{ecpsap1, ecpsap2, MM}. In \cite{ecpsap1}, a vulnerability is highlighted that allows the sender and the person with the viewing key to jointly derive the private key of the stealth address. In contrast, \cite{ecpsap2} shows that the private key of the stealth address can be determined by the sender alone without the need for collaboration. Furthermore, \cite{MM} introduces four pairing-based protocols designed to overcome the limitations identified in \cite{ecpsap1, ecpsap2}: three ECPDKSAPs (Elliptic Curve Pairing Dual-Key Stealth Address Protocols) and ECPSKSAP (Elliptic Curve Pairing Single-Key Stealth Address Protocol). Among them, only the final variant, later named the Curvy protocol, is compatible with Ethereum. Implementation results \cite{MM2} for the Curvy protocol demonstrate that its parsing process is approximately five times faster than the DKSAP implementation described in \cite{textbook1}. The Curvy protocol is currently the fastest SAP and for this reason we compare it (in the section Implementation Results) with the most efficient LWE protocol from this paper.

Most of the currently most widely used stealth address protocols are based on the DKSAP and the Zerocash protocol. Umbra Cash \cite{textbook8} and Fluidkey \cite{textbook9} are based on DKSAP, while Railgun \cite{textbook10} and Labyrinth \cite{textbook11} are based on the Zerocash protocol. The key difference is that DKSAP based protocols allows funds to be sent to a standard Ethereum address while breaking links to the recipient’s identity, the transaction remains traceable, exposing the sender's identity through the stealth address's history. In contrast, the Zerocash based protocols conceals the sender and transaction value but operates within a single contract, rather than as a standard transaction to a typical Ethereum address.

 Along with the development of the stealth address protocols based on the elliptic curve cryptography, there was a need for the development of the post-quantum stealth address protocols. Namely, National Institute of Standards and Technology (NIST) in their report \cite{nist} emphasizes the urgency of transitioning to PQC due to the potential "harvest now, decrypt later" threat. It has released three PQC standards - lattice-based and hash-based methods, as initial steps toward this transition. Code-based, isogeny-based and multivariate protocols were also proposed, but those proposals were rejected due to insufficient security and/or efficiency and key size. Migration, expected to span 10–20 years, involves significant challenges, including the integration of PQC algorithms into various systems. Also note that \textit{National Security Memorandum 10} (NSM-10) targets 2035 for Federal systems to adopt PQC, although timelines may vary depending on system complexity and risk.

So, quantum computing poses a theoretical risk to current encryption methods, including blockchain security. Therefore, Ethereum’s future security relies on integrating post-quantum cryptography to mitigate threats posed by quantum computers. The shift is crucial to protect assets, smart contracts, and user data from vulnerabilities in current public key cryptography. Ethereum’s roadmap, led by Vitalik Buterin, includes “The Splurge” \cite{textbook5},  a phase focused on adding protections against future quantum computing risks to keep the blockchain secure. 

To date, limited effort has been devoted to the development of post-quantum stealth address protocols. For instance, the protocol presented in \cite{fhedksap} incorporates fully homomorphic encryption \cite{fhe}, which is a recognized post-quantum technique. However, the core protocol is essentially a variant of DKSAP that relies on elliptic curve cryptography. Therefore, it does not qualify as post-quantum secure, despite the claims of the authors. Furthermore, \cite{textbook1} highlights that lattice-based stealth addresses represent an area requiring further research.

Motivated by these papers, we propose the construction of lattice-based stealth address protocols. These protocols leverage the following key encapsulation mechanisms: Frodo \cite{frodo-kem-paper} (based on LWE), NewHope \cite{newhope-paper} (based on Ring-LWE) and Kyber \cite{kyber} (based on Module-LWE).

\section{Post-Quantum hard problems}

In this section, we provide an overview of some of the foundational problems in the post-quantum cryptography, focusing on the LWE problems and its structured variants, Ring-LWE and Module-LWE. These problems not only underpin the security of many cryptographic constructions but also offer efficient implementations due to their algebraic structures. The foundational importance of lattice problems in cryptography was first highlighted by Ajtai, who demonstrated the hardness of generating lattice problems with cryptographic relevance, in his paper \cite{ajtai}. His results laid the groundwork for developing lattice-based cryptographic schemes, including those based on the LWE problem.

We begin by introducing lattices and ideal lattices, SVP and SIS, which form the mathematical basis for these problems, and then proceed to define the LWE, Ring-LWE, and Module-LWE problems, along with their applications in cryptographic protocols.

\subsection{Notation and basic definitions}

In our protocols, we use a ring of integers $\mathbb{Z}$ and the polynomial ring $R = \mathbb{Z}[x]/(x^n + 1)$, where $n$ is a power of 2. For an integer $q$, we use $\mathbb{Z}_q = \mathbb{Z}/q\mathbb{Z}$ to denote the ring of integers modulo $q$, and $R_q$ to denote $R/qR=\mathbb{Z}_q[x]/({x^n+1})$. We denote by regular letters elements in $R, R_q, 
\mathbb{Z}, \mathbb{Z}_q$, with lowercase bold letters vectors from $R_q^k$ (where $k>1$) and with uppercase bold letters matrices from $R_q^{k \times k}$ (where $k>1$).

Following definitions (see \cite{lattice5}) of a lattice and its basis, provide a formal framework for understanding lattice-based problems.

\begin{definition}
Let $\mathbb{R}^m$ be the $m$-dimensional Euclidean space. A lattice in $\mathbb{R}^m$ is the set 
$$
\mathcal{L}(\mathbf{b}_1, \dots, \mathbf{b}_n) = 
\left\{
\sum_{i=1}^n x_i \mathbf{b}_i : x_i \in \mathbb{Z}
\right\}
$$
of all integral combinations of $n$ linearly independent vectors $\mathbf{b}_1, \dots, \mathbf{b}_n$ in $\mathbb{R}^m$ (where $m \geq n$). The integers $n$ and $m$ are called the rank and dimension of the lattice, respectively. The sequence of vectors $\mathbf{b}_1, \dots, \mathbf{b}_n$ is called a lattice basis.
\end{definition}

A lattice is also characterized as an additive subgroup of $\mathbb{R}^m$ that is discrete, meaning it has no accumulation points other than infinity. This implies that the lattice consists of isolated points and forms a regular grid-like structure in the vector space. 

To define an ideal lattice, we first introduce the concept of an ideal.

\begin{definition} An ideal $\mathcal{I}$ of the ring $R$ is an additive subgroup $\mathcal{I} \subseteq R$ that is closed under multiplication by elements of $R$, meaning $v \cdot r \in \mathcal{I},$ $\forall v \in \mathcal{I}$ and $\forall r \in R$.
\end{definition}

An ideal lattice is a lattice derived from an ideal $\mathcal{I}$ in a commutative ring $R$ under a specific embedding, such as the coefficient embedding.  This multiplicative property introduces unique structural characteristics to ideal lattices.

Lattice-based cryptography relies on several hard computational problems. Among these, some are particularly relevant for their direct application in cryptographic schemes, and we will now define three of them (the definitions are taken from \cite{lattice3}, where one can also find definitions for the Decisional Approximate SVP (\(\text{GapSVP}_\gamma\)), Approximate Shortest Independent Vectors Problem (\(\text{SIVP}_\gamma\)), and Bounded Distance Decoding Problem (\(\text{BDD}_\gamma\))). 

The first is the Shortest Vector Problem (SVP), which is central to understanding lattice structures and their cryptographic utility.

\begin{definition}\textbf{(SVP)}
Given an arbitrary basis \( \mathbf{B} \) of some lattice \( \mathcal{L} = \mathcal{L}(\mathbf{B}) \), find a shortest nonzero lattice vector, i.e, a vector \( \mathbf{v} \in \mathcal{L} \) for which $$ \|\mathbf{v}\| = \min_{\mathbf{u} \in \mathcal{L} \setminus \{\mathbf{0}\}} \|\mathbf{u}\| .$$
\end{definition}

In cryptography, approximation variants of lattice problems are often more practical due to their reduced computational complexity. These problems are parameterized by an approximation factor \( \gamma \geq 1 \), which is typically a function of the lattice dimension $n$. For instance, the approximation version of SVP is known as \( \text{SVP}_\gamma \).

\begin{definition}\textbf{ (\( \text{SVP}_\gamma \))}
Given a basis \( \mathbf{B} \) of an \( n \)-dimensional lattice \( \mathcal{L} = \mathcal{L}(\mathbf{B}) \), find a nonzero vector \( \mathbf{v} \in \mathcal{L} \) for which $$\|\mathbf{v}\| \leq \gamma(n) \cdot \min_{\mathbf{u} \in \mathcal{L} \setminus \{\mathbf{0}\}} \|\mathbf{u}\|. $$ 
\end{definition}

Another crucial problem in lattice cryptography is the Short Integer Solution (SIS) problem. Informally, the SIS problem asks us to find a "short" nonzero integer combination of given lattice vectors that sums to zero. 

\begin{definition}\textbf{ (\( \text{SIS}_{n,q,\beta,m} \))}
Given \( m \) uniformly random vectors \( \mathbf{a}_i \in \mathbb{Z}_q^n \), forming the columns of a matrix \( \mathbf{A} \in \mathbb{Z}_q^{n \times m} \), find a nonzero integer vector \( \mathbf{z} \in \mathbb{Z}^m \) of norm \( \|\mathbf{z}\| \leq \beta \) such that 
\[
 \mathbf{A} \mathbf{z} = \sum_i \mathbf{a}_i \cdot z_i = 0 \in \mathbb{Z}_q^n.
\] 
\end{definition}

\subsection{LWE problems}

LWE problems are introduces by  Regev in the paper \cite{regev}. This work not only introduced the LWE problem but also demonstrated its profound implications for cryptography, linking its hardness to lattice problems.

To formalize LWE problems, we first define (see \cite{lattice3}) the LWE distribution. The LWE distribution encapsulates the idea of introducing noise into linear equations over a finite field, making it computationally hard to recover the secret vector.

\begin{definition}
\textbf{(LWE distribution)} For a vector $\textbf{s} \in \mathbb{Z}_q^n$ called the secret, the LWE distribution $A_{\textbf{s}, \chi}$ over $\mathbb{Z}_q^n \times \mathbb{Z}_q$ is sampled by choosing $\mathbf{a} \in \mathbb{Z}_q^n$ uniformly at random, choosing $e $ from $ \chi$, and outputting $\textbf{a}$ and $b= ( \langle \textbf{s}, \textbf{a} \rangle + e) \mod q.$
\end{definition}

According to Peikert (see \cite{lattice3}), the following two definitions represent the two main LWE problems: Search-LWE (to find the secret given LWE samples) and Decision-LWE (to distinguish between LWE samples and uniformly random ones).

\begin{definition}
\textbf{(Search-LWE}$_{n,q,\chi,m}$\textbf{)} Given $m$ independent samples $(\textbf{a}_i, b_i) \in \mathbb{Z}_q^n \times \mathbb{Z}_q$ from $A_{\textbf{s}, \chi}$ for a uniformly random $\textbf{s} \in \mathbb{Z}_q^n$ (fixed for all samples), find $\textbf{s}$.
\end{definition}

\begin{definition}
\textbf{(Decision-LWE}$_{n,q,\chi,m}$\textbf{)} For $m$ independent samples $(\textbf{a}_i, b_i) \in \mathbb{Z}_q^n \times \mathbb{Z}_q$, where every sample is distributed according to either: $(1) \,A_{\textbf{s}, \chi}$ for a uniformly random $\textbf{s} \in \mathbb{Z}_q^n$ (fixed for all samples), or $(2)$ the uniform distribution, distinguish which is the case (with non-negligible advantage).
\end{definition}

Peikert in \cite{lattice3} highlights a duality between the Short Integer Solution (SIS) and LWE problems, where the SIS and respective LWE lattices are dual to each other up to a scaling factor of $q$. This duality provides theoretical insights into the robustness of LWE, linking it to foundational lattice problems that are resistant to both classical and quantum attacks.
In \cite{regev}, it is proved that both Search-LWE and Decision-LWE are computationally hard problems, provided that the Shortest Vector Problem (SVP) in lattices is hard in the worst-case. Specifically, Regev's reduction shows that solving LWE efficiently would imply an efficient algorithm for solving SVP in certain structured lattices, which is widely believed to be intractable. This foundational result establishes the strong security guarantees of LWE by connecting its hardness to well-studied lattice problems, which remain difficult even for quantum computers.

\subsection{Ring-LWE problems}

A major drawback of schemes based on the LWE problem is their inefficiency in practical applications, as they require significant computation time and have excessively large key sizes. In this context, more efficient schemes can be constructed using the Ring-LWE problem.

The concept of Ring-LWE was first introduced in 2010 in the paper \cite{regev2} by  Lyubashevsky, Peikert and Regev. This work revolutionized lattice-based cryptography by leveraging the algebraic structure of polynomial rings, offering a more efficient alternative to LWE while maintaining its robust security foundations. To formally describe the Ring-LWE problem, we first define (see \cite{lattice3}) the underlying distribution and then outline the search and decision variants of the problem.

\begin{definition}
\textbf{(RLWE Distribution)} For an $s \in R_q$ called secret, the Ring-LWE distribution $A_{s, \chi}$ over $R_q \times R_q$ is sampled by choosing $a \in R_q$ uniformly at random, choosing $e $ from $ \chi,$ and outputting $a$ and $b= (s \cdot a + e) \mod q.$
\end{definition}

Just as in the case of LWE, the Ring-LWE problem is categorized into two fundamental variants: the search problem, which involves recovering the secret, and the decision problem, which involves distinguishing samples generated with a secret from uniformly random samples.

\begin{definition} \textbf{(Search-RLWE}$_{q,\chi,m}$\textbf{)} Given $m$ independent samples $(a_i, b_i) \!\in\! R_q \!\times\! R_q$
 from $A_{s, \chi}$ for a uniformly random $s \in R_q$ (fixed for all samples), find $s$.
\end{definition}

\begin{definition}
\textbf{(Decision-RLWE}$_{q,\chi,m}$\textbf{)} For $m$ independent samples $(a_i, b_i) \! \in \! R_q \! \times \! R_q$, where every sample is distributed according to either:  (1) $A_{s, \chi}$ for a uniformly random $s \in R_q$ (fixed for all samples), or (2) the uniform distribution, distinguish which is the case (with non-negligible advantage).
\end{definition}

 According to the Main Theorem 1 from \cite{regev2}, if it is hard for polynomial-time quantum algorithms to approximate the search version of the Shortest Vector Problem (SVP) in the worst case on ideal lattices within a fixed polynomial factor, then any polynomial number of samples drawn from the Ring-LWE distribution are pseudorandom to any polynomial-time (possibly quantum) adversary.

 This result establishes a strong theoretical foundation for the security of Ring-LWE by linking it to well-studied lattice problems. Building on this foundation, Ring-LWE retains the strong security guarantees of LWE, as its hardness can also be reduced to worst-case problems on ideal lattices, while achieving greater efficiency and practicality in cryptographic applications.

\subsection{Module-LWE problems}

In the paper \cite{fhe-gentry} is introduced the General Learning With Errors (GLWE) problem, for the purpose of constructing a fully homomorphic encryption scheme without bootstrapping, which was the first step towards considering the Module-LWE problem.

Module-LWE is explained in detail in the paper \cite{module}. This work extended the foundational principles of LWE and Ring-LWE to module structures, providing a versatile framework that enables the development of cryptographic schemes combining both efficiency and adaptability. As in the case of LWE and Ring-LWE problems, we define distributions, as well as the associated search and decision problems. These definitions are based on the work in \cite{module}, but they have been adapted to follow the style of writing the definitions presented in subsections 3.2 and 3.3.

\begin{definition}
\textbf{(Module-LWE Distribution)} For an $\textbf{s} \in R_q^k$ called secret, the Module-LWE distribution $A_{\textbf{s}, \chi}$ over $R_q^k \times R_q$ is sampled by choosing $\textbf{a} \in R_q^k$ uniformly at random, choosing $e $ from $ \chi,$ and outputting $\textbf{a}$ and $b = ( \langle \textbf{s}, \textbf{a} \rangle + e) \mod q.$
\end{definition}

\begin{definition} \hspace{-2.5mm}
\textbf{(Search-MLWE}$_{q,\chi,m,k}$\textbf{)} \hspace{-1.5mm} Given $m$ independent samples \mbox{$(\textbf{a}_i, b_i) \! \in \! R_q^k \! \times \! R_q$} from $A_{s, \chi}$ for a uniformly random $\textbf{s} \in R_q^k$ (fixed for all samples), find $\textbf{s}$.
\end{definition}

\begin{definition} \hspace{-2.8mm}
\textbf{(Decision-MLWE}$_{q,\chi,m,k}$\textbf{)} \hspace{-1.7mm} For $m$ independent samples \mbox{$(\textbf{a}_i, b_i) \! \in \! R_q^k \! \times \! R_q$}, where every sample is distributed according to either:  (1) $A_{\textbf{s}, \chi}$ for a uniformly random $\textbf{s} \in R_q^k$ (fixed for all samples), or (2) the uniform distribution, distinguish which is the case (with non-negligible advantage).
\end{definition}

Module-LWE generalizes Ring-LWE by working with module over a polynomial ring instead of just the rings themselves, allowing for more flexible dimensions and balancing security and efficiency. It inherits the hardness assumptions of both LWE and Ring-LWE, under the appropriate parameter settings. Its flexibility and efficiency have made it a foundation for modern PQ cryptographic schemes, such as Kyber \cite{kyber} and Dilithium \cite{dilithium}, which are finalists in the NIST PQ standardization process.

\section{Stealth Address Protocols}

\begin{figure}[H]
  \centering
    \includegraphics[scale=0.52]{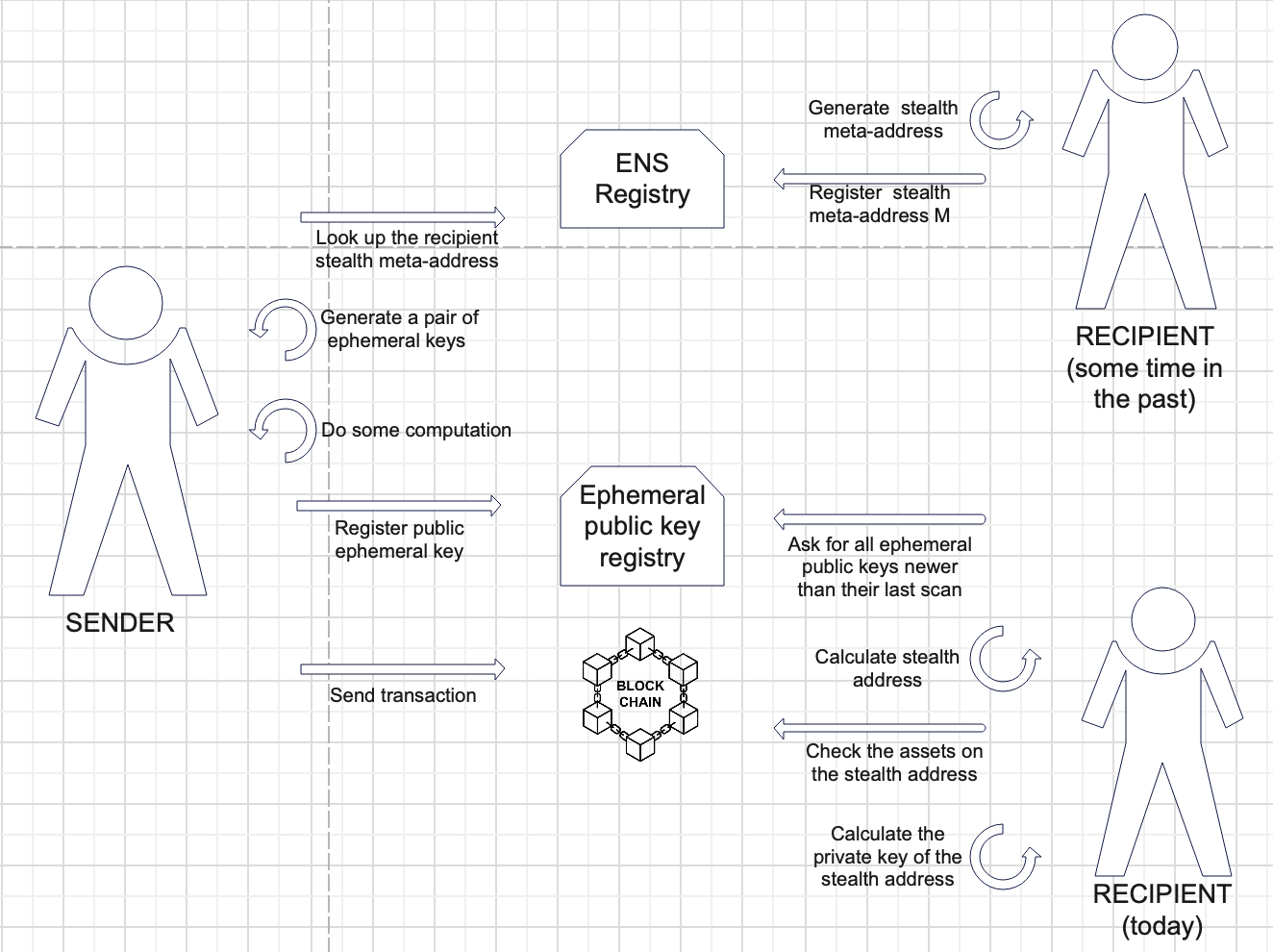}
      \caption{Stealth Address Protocol}
      \label{fig:sap}
\end{figure}

With the stealth address protocol, we aim to ensure that the recipient of a transaction remains anonymous, while it is possible for anyone to easily send assets to them using their stealth meta-address. This stealth meta-address does not link the recipient to the newly created, single-use stealth address. As previously mentioned, a stealth address protocol can be constructed using various cryptographic methods. However, these protocols have a common central concept that does not rely on a specific cryptographic technique. This concept is introduced in the following lines and shown in Figure \ref{fig:sap}.

Since the original idea related to these protocols involves an Ethereum Name Service (ENS) registry, we will briefly discuss what this registry is, while pointing out that the protocols remain functional even with minor changes or the use of a different registry. The ENS registry is a domain name system built on the Ethereum blockchain. ENS enables the conversion of randomly generated letters and numbers in a standard Ethereum address into more easily recognizable words, such as the name of a person or a brand.

The complete workflow of a stealth address scheme can be summarized as follows:

\begin{itemize}
\item[1)] The recipient generates a key pair or pairs (private and public key) for the stealth meta-address.
\item[2)] The recipient registers the stealth meta-address by adding an ENS record containing the public key(s).
\item[3)] The sender retrieves the recipient's stealth meta-address from the ENS registry.
\item[4)] The sender generates an ephemeral key pair (private and public key).
\item[5)] Using a cryptographic algorithm, the sender combines their ephemeral private key with the recipient's meta-address to compute the recipient's stealth address.
\item[6)] The sender publishes their ephemeral public key to the ephemeral public key registry.
\item[7)] The sender sends assets to the recipient's address.
\item[8)] To discover stealth address belonging to them, the recipient scans all new ephemeral public keys in the registry since their last scan.
\item[9)] The recipient computes their new stealth address.
\item[10)] The recipient derives the private key that corresponds to the new stealth address.
\end{itemize}

Newer stealth address protocols introduce the concept of a view tag, which simplifies the recipient's search in the ephemeral registry. Additionally, some protocols require the recipient to generate two key pairs. These differences are discussed in the following subsections.

\subsection{View tag}

The Monero blockchain was the first to implement the use of a view tag. Adding a view tag to the ephemeral public key is a technique designed to improve the efficiency of searching the ephemeral public key registry. This approach eliminates the need to compute all the steps for each ephemeral public key and instead focuses only on the keys that the view tag matches.

In some protocols, the view tag should not be larger than 1 byte, otherwise the recipient of the transaction could potentially be de-anonymized. However, in the ECPDKSAP and LWE protocols presented in this paper, the view tag can be larger. This enables a much more efficient search for the recipient in the registry.

Therefore, in the previously mentioned scheme, the only modification would occur in step 6, where the sender includes the view tag to the ephemeral public key registry along with the ephemeral public key.

\subsection{Viewing key}

The concept of a viewing key was introduced in 2014 with the development of DKSAP. This key allows a recipient to share access with a third party, such as a tax inspector, enabling them to view all stealth transactions associated with the recipient. Importantly, possessing the viewing key does not grant the ability to spend assets linked to the stealth addresses. This limitation arises because the viewing key holder can only derive the public keys of the stealth addresses, while the private keys, necessary for spending, remain inaccessible.

From a regulatory perspective, the viewing key plays a crucial role in stealth address protocols. It strikes a balance between the privacy of the recipient and the need for traceability of transactions when required. This feature is particularly valuable in applications that involve decentralized identities and verifiable credentials.

For protocols that include a viewing key (such as DKSAP, ECPDKSAPs and LWE SAPs, discussed in this paper), the first step of the scheme is to generate two pairs of keys (private and public). One of these pairs includes the private viewing key. Therefore, the first step in such protocols may require generating one more key pairs, depending on the specific protocol requirements.

\section{LWE Stealth Address Protocols}

In this section, we provide a general overview of how our LWE protocols work. The centerpiece of this section is the MLWE SAP, as it is our most performant LWE SAP version. But before we get into the protocols, we provide the necessary background and introduce the notation for the terms we will use in this section.

\subsection{Background}

We begin this section with the modular reduction and define infinity norms in the corresponding spaces, which we will use in the rest of the paper.

\begin{definition}\textbf{\mbox{(Modular reduction)}}
    For any positive integer $q$ and $r \in \mathbb{Z}$, we define $r' := r \bmod{q}$, $ r'\in \mathbb{Z}_q$, to be a unique element such that $0 \leq r' < q$.
\end{definition}

\begin{definition}\textbf{(Symmetric modular reduction)}
For an even positive integer $q$ and $r \in \mathbb{Z}$, we define 
$$\displaystyle
r' := r \bmod^+{q} =
\begin{cases} 
r & \text{if } r \leq \frac{q}{2}, \\
r-q   & \text{if } r > \frac{q}{2} \\
\end{cases} 
.$$
\end{definition}

For odd $q$, we would have a case that $\frac{q-1}{2} \leq r' < \frac{q-1}{2}$, everything else stays the same. 
Note that newly defined $r'$ belongs to $\mathbb{Z}_q$.

For an element $r \in \mathbb{Z}_q$, we define $\lVert r \rVert_\infty := \lvert r \bmod^+{q} \rvert$. For $r \in R_q$ we define it as $\lVert r \rVert_\infty := \max\limits_{i} \lVert r_i \rVert_\infty$, where $r_i$ are coefficients of polynomial $r$ in $R_q$, i.e.\ $r_i \in \mathbb{Z}_q$. For $\textbf{r} = (r_1, r_2, ..., r_k)  \in R_q^k$, we define it as $\lVert \mathbf{r} \rVert_\infty := \max\limits_{j} \lVert r_j \rVert_\infty$, where $r_j \in R_q$.

In the following, we introduce some definitions that are necessary due to the concepts we use in the rest of the paper and their notation.

\begin{definition}\textbf{(Ceiling function)}  
Let $x \in \mathbb{Q}$ be a rational number. The ceiling function, denoted by $\lceil x \rceil$, maps $x$ to the smallest integer greater than or equal to $x$, is defined as  
\[ 
\lceil x \rceil := \min \{ z \in \mathbb{Z} \mid z \geq x \}.  
\]  
\end{definition}  

\begin{definition}\textbf{(Rounding to the nearest integer)}
Let $x \in \mathbb{Q}$ be a rational number. The rounding to the nearest integer function, denoted by $\lceil x \rfloor$, maps $x$ to the closest integer to $x$, with the exception that if x is exactly between two integers, we map it to the greater integer, is defined as
$$\lceil x \rfloor := \min{\{z \in \mathbb{Z} \mid z > x - 0.5\}}.$$
\end{definition}

\begin{definition}\textbf{(Centered binomial distribution)}  
Let $\eta$ be a positive integer. The centered binomial distribution $B_\eta$ is defined as the probability distribution over  
\[ 
x = \sum_{i=1}^\eta (a_i - b_i),  
\]  
where $a_i, b_i \sim \text{Ber}(0.5)$, for $i \in \{1, 2, \dots, \eta\}$, are independent random variables sampled from the Bernoulli distribution with parameter $p = 0.5$.  
\end{definition}  

Note that the output $x$ belongs to the set $\{-\eta, -\eta+1, \dots, \eta-1, \eta\}$.

In our protocols, we use polynomials (and vectors of polynomials, as well as matrices of polynomials) from $R_q$ and $B_\eta$, where $B_\eta$ is a centered binomial distribution with parameter $\eta$. In the Kyber \cite{kyber} key encapsulation mechanism, a centered binomial distribution $B_\eta$ is used for $\eta=2$ and $\eta=3$.

When we write that a polynomial is from $B_\eta$, this means that each coefficient is sampled from $B_\eta$.  A vector of polynomials from $R^k$ can be sampled from $B^k_\eta$.

Extendable Output Function ($\text{XOF}$) is a function on bit strings where the output can be extended to any desired length. If we want $\text{XOF}$ to take $x$ as input and then produce a value $y$ that is distributed uniformly over a set $T$, we write as $y \sim T : = \text{XOF}(x)$. This function is deterministic, i.e.\ it always generates the same $y$ for a given $x$. If we write $\textbf{A} \sim R_q^{k \times k} := \text{XOF}(x)$, we mean that the output of XOF is $\textbf{A} \in R_q^{k \times k}$.

Compression and decompression functions are used for optimization to reduce the parameters. They discard some of the low-order bits of the Kyber public key and its ciphertexts.

\begin{definition}\textbf{(Compression)}
 Function 
 $\text{Compress}_q(x, d)$, where $x \in Z_q$, $d$ and $q$ are positive integers,  such that $d < \lceil \log_2 q\rceil$, outputs an integer from the set $\{0, 1, ..., 2^d - 1\}$ in the following way
$$
\text{Compress}_q(x, d) := \left\lceil \frac{2^d \cdot x}{q} \right\rfloor \bmod{2^d}.
$$
\end{definition}

\begin{definition}\textbf{(Decompression)} Function $\text{Decompress}_q(x, d)$, where $x \in Z_{2^d}$, $d$ and $q$ are positive integers, such that $d < \lceil \log_2 q\rceil$, outputs an integer from the set $\{0, 1, ..., q-1\}$ in the following way
$$
\text{Decompress}_q(x, d) := \left\lceil \frac{q\cdot x}{2^d} \right\rfloor \bmod{q} .
$$
\end{definition}

Compression and decompression have a certain impact on the correctness of decryption, which is the subject of the following theorem.

\begin{theorem} Let $x \in \mathbb{Z}_{q}$, $d$ and $q$ be positive integers, where $d < \lceil \log_2 q\rceil$. Let 
$$x' := \text{Decompress}_q(\text{Compress}_q(x,d),d).$$
Then $x'$ is close to $x$, i.e.\ $\lvert (x' - x) \bmod^+{q} \rvert \leq \left\lceil \dfrac{q}{2^{d+1}} \right\rfloor.$
\end{theorem}

We can extend these functions to $R_q$ and $R_q^k$ by applying them to each coefficient of the polynomials in $R_q$ and $R_q^k$. The theory behind compression and decompression comes mostly from the Kyber paper \cite{kyber-original}.

\subsection{Kyber}

Kyber is a secure post-quantum Key Encapsulation Mechanism (KEM) based on the Module-LWE problem. It is one of the finalists in the NIST competition for post-quantum cryptography.

There is a Kyber512, a Kyber768 and a Kyber1024 version, each with different security levels. Kyber512 has a security level roughly equivalent to AES128, Kyber768 is roughly equivalent to AES192 and Kyber1024 is roughly equivalent to AES256.
We use Kyber for secret sharing in our MLWE SAP.

\subsubsection{Kyber's IND-CPA-secure encryption}

By IND-CPA (Indistinguishability under Chosen Plaintext Attack) we mean the property where an adversary, even if he has access to an encryption oracle, cannot distinguish between the ciphertexts of any two arbitrarily chosen plaintext messages \cite{katz-lindell}. For a formal treatment of Kyber IND-CPA security, we refer the reader to the original paper \cite{kyber-original}.

CCA (Chosen Ciphertext Attack) security is a stronger definition of security than IND-CPA in terms of power of the adversary. In this scenario, the adversary has access to a decryption oracle as well as an encryption oracle \cite{katz-lindell}.

Let $M = \{0,1\}^{256}$ be the message space, where each $m \in M$ can be viewed as a polynomial with coefficients from the set \{0, 1\}. We have the parameters $n, k, q, d_t , d_u$ and $d_v$. For all versions of Kyber, the parameter $n$ is 256, where $n-1$ is the maximum degree of the polynomials of $R_q$, where $q=3329$. The parameter $k$ represents the dimension of the module over a polynomial ring. For Kyber512 it is $k=2$, for Kyber768 $k=3$ and for Kyber1024 $k=4$. The parameters $d_u, d_v$ and $d_t$ are used for compression and decompression functions. In the case of Kyber512 and Kyber768 $d_t = d_u = 10$, $d_v=4$ and in the case of Kyber1024, $d_t = d_u = 11$, $d_v = 5$. 

In this section we define the PKE (Public-Key Encryption) scheme, which is a building block of the CCA secure key encapsulation mechanism.
A PKE scheme is a triple of algorithms (KeyGen, Enc, Dec).

KeyGen generates a pair that contains a private and a public key $(sk, pk)$.
\begin{algorithm}[H]
\caption{Kyber.CPA.KeyGen(): key generation}
\begin{algorithmic}[1] 
\State \( \rho, \sigma \gets \{0,1\}^{256} \)
\State $\textbf{A} \sim R_q^{k \times k} : = \text{XOF}(\rho)$ 
\State $(\textbf{s}, \textbf{e}) \sim B^k_{\eta} \times B^k_\eta : = \text{XOF}(\sigma)$
\State $\textbf{t} : = \text{Compress}_q(\textbf{A} \textbf{s} + \textbf{e}, d_t)$ 
\State \Return $(pk : = (\textbf{t}, \rho), sk : = \textbf{s})$
\end{algorithmic}
\end{algorithm}

Enc is an encryption algorithm that uses a public key and randomness to securely encrypt a message. The inputs to the algorithm are the public key $pk$, the message $m$ and a value $l$ (if not specified, it is by default a uniformly random value). The output is the encryption of the message $m$ in the form $(\textbf{u}, v) \in \left(\{0, 1\}^{256\cdot d_u}\right)^k \times \{0, 1\}^{256 \cdot d_v}$.

\begin{algorithm}[H]
\caption{Kyber.CPA.Enc($pk = (\textbf{t}, \rho), m \in M, l \in \{0, 1\}^{256})$: encryption}
\begin{algorithmic}[1] 
\If{$l = null$}
    \State $l \gets \{0, 1\}^{256}$
\EndIf
\State $\textbf{t} : = \text{Decompress}_q(\textbf{t}, d_t)$
\State $\textbf{A} \sim R_q^{k \times k} : = \text{XOF}(\rho)$
\State $(\textbf{r}, \textbf{e}_1, e_2) \sim B_\eta^k \times B_\eta^k \times B_\eta : = \text{XOF}(l)$
\State $\textbf{u} : = \text{Compress}_q(\textbf{A}^T \textbf{r} + \textbf{e}_1, d_u)$ 
\State $v : = \text{Compress}_q(\textbf{t}^T\textbf{r} + e_2 + \lceil q/2 \rfloor \cdot m, d_v)$
\State \Return $c : = (\textbf{u}, v)$
\end{algorithmic}
\end{algorithm}

Dec is a decryption algorithm that receives the private key $sk$ and the ciphertext $c$ as inputs and outputs a decryption of $c$.

\begin{algorithm}[H]
\caption{Kyber.CPA.Dec($sk = \textbf{s}, c = (\textbf{u}, v)$): decryption}
\begin{algorithmic}[1] 
\State $\textbf{u} : = \text{Decompress}_q(\textbf{u}, d_u)$ 
\State $v : = \text{Decompress}_q(v, d_v)$
\State \Return $\text{Compress}_q(v - \textbf{s}^T\textbf{u}, 1)$
\end{algorithmic}
\end{algorithm}

Compression is used in Algorithm 3 to decrypt to $1$ if $v-\textbf{s}^T\textbf{u}$ is closer to $\lceil q/2 \rfloor$ than to $0$, and to decrypt to $0$ otherwise. The algorithms are taken from \cite{kyber-original}.

The decryption function of Kyber PKE is correct with a certain probability and we define this correctness in the next definition.

\begin{definition}\textbf{(PKE Correctness \cite{delta-correctness-paper})}
PKE is $(1-\delta)$-correct if 
$$\textbf{E}[\min\limits_{m\in M} Pr[\text{Dec}(sk, \text{Enc}(pk, m)] = m] \geq \delta,$$ 
where the mathematical expectation $\textbf{E}$ is taken over $(pk, sk) \gets \text{KeyGen()}$.
\end{definition}

In the next theorem we want to quantify the probability that Kyber PKE decrypts correctly.

\begin{theorem}
    Let $k, d_t, d_u$ and $d_v$ be positive integers. Let $\textbf{s}, \textbf{e}, \textbf{r}, \textbf{e}_1 \in B^k_\eta, e_2 \in B_\eta$  be random variables. Also let $\textbf{c}_t \in \psi^k_{d_t}, \textbf{c}_u \in \psi^k_{d_u}, c_v \in \psi_{d_v}$ be distributed according to the distribution $\psi^k_d$ defined as follows:
    
    1. choose uniformly random $\textbf{y} \in R_q^k;$
    
    2. \textbf{return} $(\textbf{y} -\text{Decompress}_q(\text{Compress}_q(\textbf{y},d),d)) \, mod\, ^+{q}.$ 
    
    Then Kyber.CPA is $(1 - \delta)$-correct, where \[
        \delta = \Pr\left[ \| \mathbf{e}^T \mathbf{r} + e_2 + c_v - \mathbf{s}^T \mathbf{e}_1 + \mathbf{c}_t^T \mathbf{r} - \mathbf{s}^T \mathbf{c}_u \|_\infty \geq \lfloor q / 4 \rfloor \right].
    \]
\end{theorem}

The parameters for all Kyber PKE are chosen so that $\delta < 2^{-128}$. We refer the reader to the proof in the original paper \cite{kyber-original}.

\subsubsection{CCA-secure KEM}

Let $G: \{0, 1\}^* \to \left(\{0,1\}^{256}\right)^2$ and $H: \{0, 1\}^* \to \{0,1\}^{256}$ be hash functions, where $\{0, 1\}^*$ represents an input of arbitrary length. The CCA-secure KEM is obtained by applying the Fujisaki-Okamoto transform to the Kyber.CPA. The Kyber.CCA.KeyGen function is similar to the Kyber.CPA.KeyGen function, but additionally generates and returns the private random key $z \in \{0,1\}^{256}$.

Encaps is an encapsulation algorithm that takes as input a public key $pk$ and outputs a ciphertext $c$ and a shared secret $K$ (a secret that in our protocol can only be calculated by the sender and the person holding the viewing key), which is the output of a hash function.

\begin{algorithm}[H]
\caption{Kyber.CCA.Encaps($pk = (\textbf{t}, \rho)$): encryption}
\begin{algorithmic}[1] 
\State $m \gets \{0, 1\}^{256}$
\State $(\hat{K}, l) := \text{G}(\text{H}(pk), m)$
\State $c = (\textbf{u}, v) := \text{Kyber.CPA.Enc}((\textbf{t}, \rho), m; l)$
\State $K :=  \text{H}(\hat{K}, \text{H}(c))$
\State \Return $(c, K)$
\end{algorithmic}
\end{algorithm}

Decaps is a decapsulation algorithm that takes as inputs the private key $\textbf{s}$, the private random key $z$, the public key $pk$ and the ciphertext $c$ output by the encapsulation and returns a shared secret $K$ if the encryption of $m'$ (which we obtain by decrypting the ciphertext $c$) is the same as the input ciphertext $c$; otherwise it returns a pseudo-random key $K : = \text{H}(z, c)$.

\begin{algorithm}[H]
\caption{Kyber.CCA.Decaps($\textbf{s}, z, pk=(\textbf{t}, \rho), c = (\textbf{u}, v)$): decryption}
\begin{algorithmic}[1] 
\State $m' : = \text{Kyber.CPA.Dec}(\textbf{s}, (\textbf{u}, v))$
\State $(\hat{K}', l') : = \text{G}(\text{H}(pk), m')$
\State $(\textbf{u}', v') : = \text{Kyber.CPA.Enc}((\textbf{t}, \rho), m'; l')$
\If{$(\textbf{u}', v') = (\textbf{u}, v)$}
    \State $K := \text{H}(\hat{K}', \text{H}(c))$
\Else
    \State $K := \text{H}(z, \text{H}(c))$
\EndIf
\State \Return $K$
\end{algorithmic}
\end{algorithm}

The algorithms are taken from \cite{kyber-original}. In the following lines, we will define a general correctness of KEM.

\begin{definition}\textbf{(KEM Correctness \cite{delta-correctness-paper})} KEM is $(1- \delta)$ correct, if 
$$Pr\left[\text{Decaps}(\text{sk}, c) = K \,\middle|\, (\text{pk}, \text{sk}) \gets \text{KeyGen()}; (K, c) \gets \text{Encaps}(\text{pk})\right] \geq \delta.$$ 
\end{definition}

If Kyber.CPA is $(1-\delta)$-correct and $G$ is a random oracle, then Kyber.CCA is also $(1 - \delta)$-correct. In the case of Kyber512 KEM $\delta < 2^{-139}$, for Kyber768 KEM $\delta < 2^{-164}$ and for Kyber1024 KEM $\delta < 2^{-174}$.

\subsection{Module-LWE Stealth Address Protocol}

Module-LWE strikes a balance between the flexibility of LWE and the efficiency of Ring-LWE, making it ideal for practical cryptographic protocols. Its structured design enables better key size and performance optimization, as demonstrated by NIST-standardized schemes such as Kyber. By utilizing Module-LWE, our stealth address protocol achieves robust security, computational efficiency and scalability.

\begin{figure}[H]
    \centering
    \includegraphics[scale=0.52]{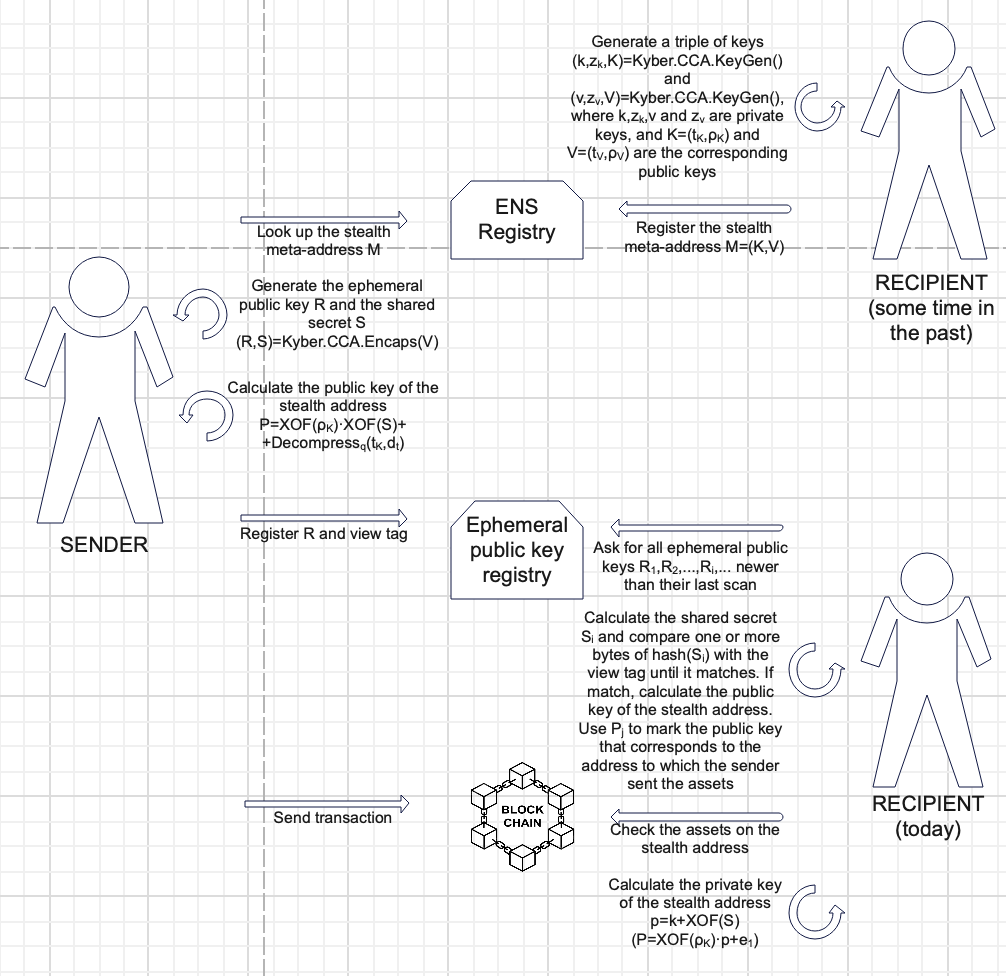}
    \caption{Module-LWE Stealth Address Protocol}
    \label{fig:protocol}
\end{figure}

Figure \ref{fig:protocol} illustrates Module-LWE SAP, which works as follows:

\begin{itemize}
\item[1)]The recipient obtains a triple of private and public keys $(k, z_k, K)$ by calling the function Kyber.CCA.KeyGen and obtains a triple of keys $(v, z_v, V)$ in the same way. Note that the public keys $K$ and $V$ consist of two components, i.e.\ $K=(t_K, \rho_K)$ and $V=(t_V, \rho_V)$.
\item[2)] The recipient registers the stealth meta-address by adding an ENS record containing the public keys $M = (K,V)$.
\item[3)] The sender retrieves the recipient's stealth meta-address from the ENS registry.
\item[4)] The sender runs the encapsulation by calling the function Kyber.CCA.Encaps with the public key $V$ as input and derives the shared secret $S$ and the ephemeral public key $R$ from it, i.e.
$$
(R,S)=\text{Kyber.CCA.Encaps}(V).
$$
Note that the sender uses a private randomness $l$ generated in the Kyber.CPA.Enc function through the Kyber.CCA.Encaps function.
\item[5)] The sender calculates the public key of the recipient's stealth address as 
$$P =\text{XOF}(\rho_K) \cdot \text{XOF}(S) + \text{Decompress}_q(t_K, d_t).$$
\item[6)] The sender publishes their ephemeral public key $R$ to the ephemeral public key registry as well as a view tag (one or more bytes of the hash of $S$).
\item[7)] The sender sends assets to the recipent's stealth address.
\item[8)] To discover the stealth address belonging to them, the recipient scans all new ephemeral public keys in the ephemeral public key registry since their last scan. Recipient for each retrieved value $R_i$ calculates the shared secret by calling Kyber.CCA.Decaps function as

$$
S_i = \text{Kyber.CCA.Decaps}(v,z_v,V,R_i).
$$

After that, the recipient compares one or more bytes of the hash of $S_i$ with the view tag until it matches.
\item[9)] Let's use $R_j$ to mark the ephemeral public keys for which the view tag matches. The recipient calculates the public keys $P_j$ in the same way as the sender. Then the recipient calculates the address and checks if it is the stealth address to which the sender sent the assets.

\item[10)] The recipient derives the private key that corresponds to the new stealth address. The private key $p$ of the stealth address is
$$p = k + \text{XOF}(S)$$
and can only be calculated by the recipient, as this requires the private key $k$, which only the recipient has. \end{itemize}

Note that the recipent's public key $P$ of the stealth address can only be calculated by the sender (who has the private randomness $l$) and the person who holds the viewing key $v$ (recipient and, for example, a tax inspector). Also note that the private key $p$ of the stealth address corresponds to the public key $P$ of the stealth address, because it is
$$P = \text{XOF}(\rho_K) \cdot \text{XOF}(S) + \text{Decompress}_q(t_K, d_t) == \text{XOF}(\rho_K) \cdot p + e_1,$$
where $e_1$ can be calculated by the recipient as $$e_1 = \text{Decompress}_q(t_K, d_t) - \text{XOF}(\rho_K) \cdot k.$$

\subsection{Ring-LWE Stealth Address Protocol}

The Ring-LWE version of our stealth address protocol uses NewHope \cite{newhope-paper} as the underlying KEM. The security of this protocol is grounded in the hardness of the Search-RLWE problem (see Definition 10 in subsection 3.3). In particular, the Search-RLWE$_{q,\chi,m}$ problem involves recovering a secret $k \in R_q$ given $m$ samples drawn from the Ring-LWE distribution \(A_{k, \chi}\) (see Definition 9 in subsection 3.3).

The main idea is to use NewHope's PKE as a building block for NewHope KEM. Instead of Kyber's KeyGen, Encaps and Decaps functions, we use NewHope's corresponding KeyGen, Encaps and Decaps functions.

Given a shared secret $S$, the recipient calculates the public key of the stealth address as follows
\[
P = A(k + \text{XOF}(S)) + e_1,
\]
where $A \in R_q$ is a public value, $k \in R_q$ is the private key of the recipient's stealth meta-address, and  $e_1 \in R_q$, which represents the added noise, is the polynomial whose coefficients are sampled from the $B_\eta$ distribution (see Definition 19). The private key of the stealth address is $$p=k+XOF(S).$$  This construction ensures that the private key $p$ cannot be recovered from the public values and thus provides security for the recipient.

\subsection{LWE Stealth Address Protocol}

The LWE version of our stealth address protocol employs FrodoKEM \cite{frodo-kem-paper}, leveraging the same principles as Kyber and NewHope. The protocol's security builds upon the foundational LWE problem, characterized as Search-LWE$_{n,q,\chi,m}$ (see Definition 7 in subsection 3.2). This problem requires recovering a secret $k \in \mathbb{Z}_q^n$ given $m$ samples from the LWE distribution \(A_{k, \chi}\) (see Definition 6 in subsection 3.2).

FrodoKEM relies on FrodoPKE for the implementation of the KeyGen, Encaps and Decaps functions required in the protocol.

Similar to the Ring-LWE protocol, the recipients calculates the public key of the stealth address as follows
\[
P = A(k + \text{XOF}(S)) + e_1,
\]
where $A \in \mathbb{Z}_q^{k \times k}$ is a public value, $k \in \mathbb{Z}_q^k$ is the private key of the recipient's stealth meta-address, and sample $e_1 \in \mathbb{Z}_q^k$ from $B_\eta^k$ represents added noise. The private key of the stealth address is $$p=k+XOF(S).$$ Unlike Module-LWE and Ring-LWE SAPs, the LWE SAP does not use a polynomial ring structure. This leads to a larger key size, but is based on simpler underlying mathematics. However, LWE SAP is the slowest of these protocols as it has a lower time efficiency due to the lack of optimizations from the polynomial ring structure.

\section{Implementation results}

In our implementation of the LWE, RLWE and MLWE stealth address protocols, we used the programming language \textit{Rust} together with the libraries KyberKEM \cite{kyber-kem-repo}, NewHope \cite{newhope-repo} and FrodoKEM \cite{frodo-kem-repo}. The complete implementations of our three PQ stealth address protocols are available in the GitHub repository \cite{pq-sap-repo}.

Since the most critical aspect to optimize in stealth address protocols is the recipient's scanning of announcements (ephemeral public keys) within the ephemeral public key registry, subsection 6.1 focuses on the comparison of this specific component in our three PQ stealth address protocols. We also compare the scan time of the ephemeral public key registry in the MLWE SAP in three variants:
\begin{itemize}
 \item without view tag;
 \item view tag is one byte of hash($S$);
 \item view tag is full hash($S$),
\end{itemize}
where $S$ is the shared secret and the hash function used is SHA-256. In subsection 6.2, we extend the analysis by comparing the scan time of the ephemeral public key registry using our most efficient protocol from this paper (MLWE SAP) with the current fastest non-PQ protocols, specifically ECPDKSAP (Curvy protocol). In subsection 6.3 we compare the scan time of the ephemeral public key registry for different KEM parameters of Kyber, NewHope and Frodo. The experiments were performed on \textit{Macbook Air with M2 (8-core) chip and 8GB RAM}.
  
\subsection{Comparison of LWE, RLWE and MLWE SAPs}

In this experiment, as in the experiments in subsections 6.2 and 6.3, 10 seeds were randomly selected to generate private keys and announcements to ensure consistency. The average execution time was measured and used for comparison.

\begin{figure}[H]
  \centering
    \includegraphics[scale=0.48]{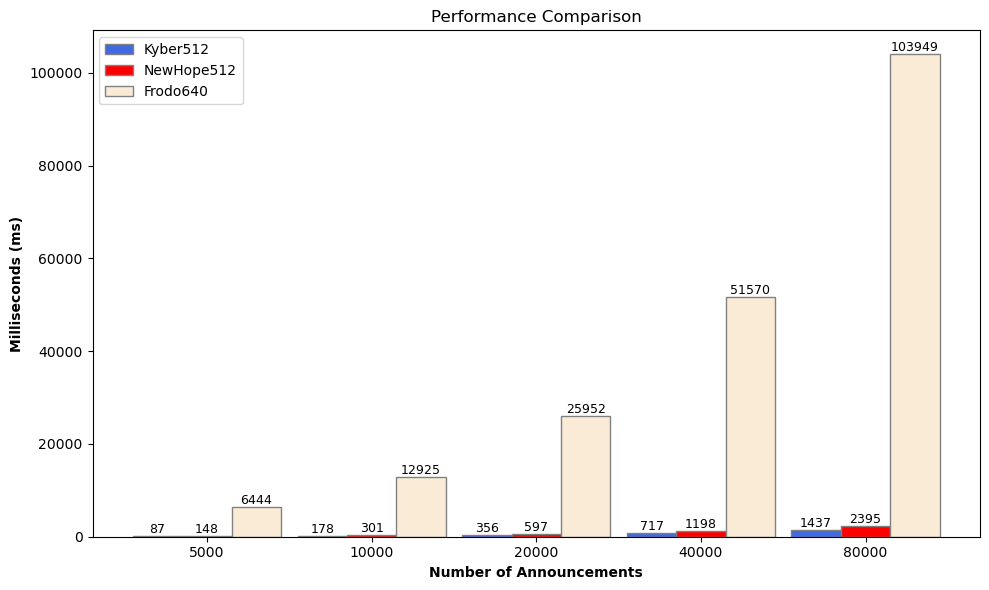}
      \caption{Performance comparison of Kyber512, NewHope512 and Frodo640}
      \label{fig:kem_comparison}
\end{figure}

All three algorithms (Kyber512, NewHope512, Frodo640) have the same security level, which corresponds approximately to AES128. For this security level, we have the best performance with sufficient security guarantees, similar to the choice of BN254 in \cite{MM}.

Figure \ref{fig:kem_comparison} shows a comparison of scan times for the ephemeral public key registry with different numbers of announcements (5000, 10000, 20000, 40000 and 80000) for the LWE, RLWE and MLWE protocols. In all scenarios, the view tag is set to one byte of $\mathrm{hash}(S)$, where $S$ is the shared secret.

Note that the LWE protocol gives the worst results for all numbers of announcements, while the MLWE protocol gives the best results for all measurements. Specifically, for 80000 announcements, MLWE SAP outperforms RLWE SAP by 40\% and LWE SAP by 98.6\%. Since MLWE SAP is our best performing protocol, we will only analyze its performance in the rest of this section.

\begin{figure}[H]
  \centering
    \includegraphics[scale=0.4]{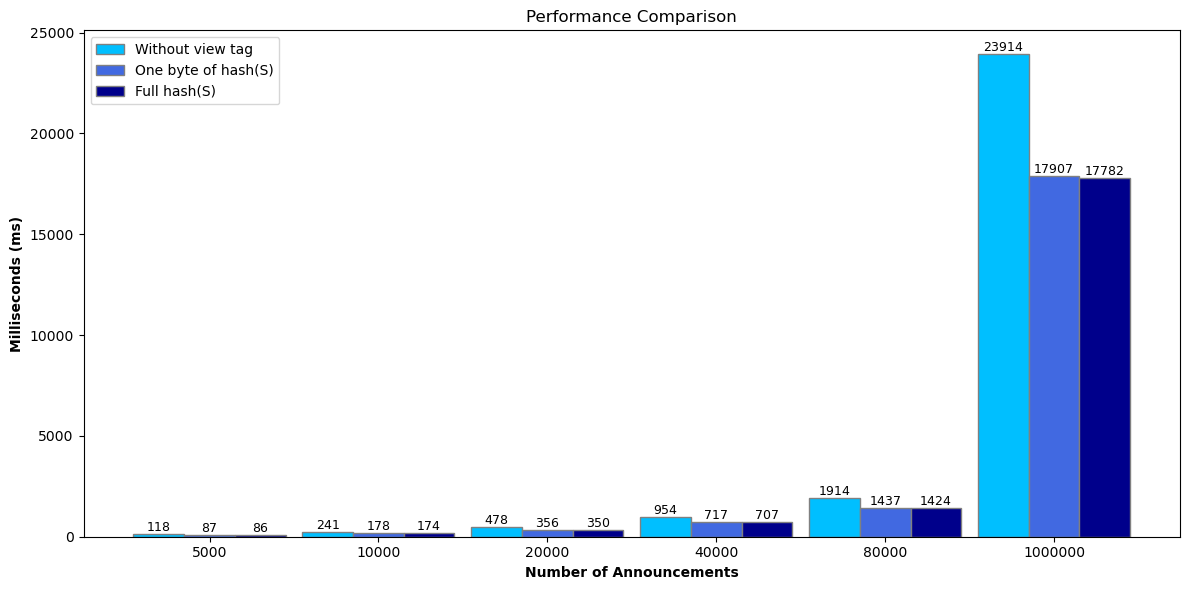}
      \caption{Module-LWE SAP - Performance comparison between the full hash($S$) view tag, one byte of hash($S$) and without view tag}
      \label{fig:comparison_full_vs_byte}
\end{figure}

In all our PQ SAPs, the view tag can be of any size by using bytes of $\mathrm{hash}(S)$ without compromising security. The reason for this is that the shared secret is used directly to calculate the public key of the stealth address and not its hash. Therefore, the full $\mathrm{hash}(S)$ can also be used as a view tag for optimization purposes. We measured how much improvement MLWE SAP can achieve by increasing the view tag to the full hash($S$). We also looked at the MLWE SAP version without the view tag. The comparison of the scan times of the ephemeral public key registry with a different number of announcements (5000, 10000, 20000, 40000, 80000 and 1000000) is shown in Figure \ref{fig:comparison_full_vs_byte}. A view tag of one byte brings a considerable acceleration compared to the version without a view tag (an improvement of 25.12\% in the case of 1000000 announcements). Increasing the view tag to the entire hash($S$) brings only a small additional acceleration (0.7\% in the case of 1000000 announcements) and requires considerably more memory space. Due to the trade-off between the efficiency of the scan time and the required storage space, a view tag of one byte is therefore the best choice.

\subsection{Comparison of MLWE SAP with Curvy protocol}

\begin{figure}[H]
  \centering
    \includegraphics[scale=0.48]{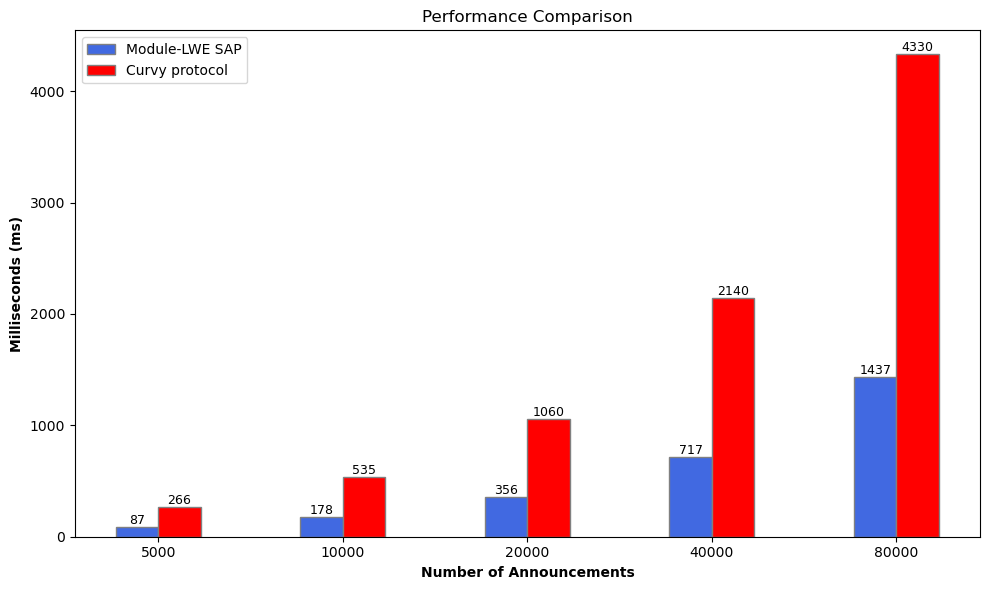}
      \caption{Performance comparison between MLWE SAP and Curvy protocol}
      \label{fig:mlwe_vs_ecc_plot}
\end{figure}

Figure \ref{fig:mlwe_vs_ecc_plot} shows a comparison of the scan times of the ephemeral public key registry for different numbers of announcements (5000, 10000, 20000, 40000 and 80000) of our implementation of MLWE SAP with the implementation of Protocol 3 (Curvy protocol) from the paper \cite{MM} (currently the most efficient SAP). The size of the view tag is 1 byte for both protocols.

The implementation of MLWE SAP shows a significant improvement compared to the mentioned implementation of the Curvy protocol (Protocol 3 from the paper \cite{MM} - the best performing protocol from this paper). For each number of announcements (5000, 10000, 20000, 40000 and 80000) the scan time of the ephemeral public key registry was reduced by about three times, especially in the case of 80000 announcements we achieved an improvement of 66.8\%.

Note that the MLWE SAP provides post-quantum security, while the Curvy protocol does not (since the security relies on the EC discrete logarithm problem), but also provides better efficiency than the Curvy protocol. However, the MLWE SAP is not Ethereum-friendly and its implementation therefore requires Ethereum to become a PQ blockchain.

\subsection{Comparison of different KEM parameters}

In subsection 6.1, we compared Kyber512, NewHope512 and Frodo 640, which correspond to the security level of AES128. There are other variants of these algorithms. The difference between these algorithms lies in the chosen parameters and their security levels. That is, Kyber768 and Frodo976 correspond to AES192 (NewHope does not have a version that corresponds to AES192), and Kyber1024, NewHope1024 and Frodo1344 correspond to AES256 in terms of security.

Table \ref{tab:kem_parameter_sets_table} illustrates a comparison of different parameter values for different KEMs using one byte of $\mathrm{hash}(S)$ as view tag for 5000 announcements.

In the most secure setting, Kyber1024 showed an improvement of 30\% and 99\% compared to NewHope1024 and Frodo1344 respectively. For Kyber768 and Frodo976, this improvement was again 99\%.

\begin{table}[H]
\centering
\begin{tabular}{|c|c|}
\hline
\textbf{Time (ms)} & \textbf{Paramset} \\
\hline
87 & Kyber512 \\
\hline
141 & Kyber768 \\
\hline
205 & Kyber1024 \\
\hline
148 & NewHope512 \\
\hline
293 & NewHope1024 \\
\hline
6444 & Frodo640 \\
\hline
13975 & Frodo976 \\
\hline
24806 & Frodo1344 \\
\hline
\end{tabular}
\caption{Time for different KEM parameter sets}
\label{tab:kem_parameter_sets_table}
\end{table}

\section{Conclusion}

The main goal of this paper is to obscure the connection between the recipient and their address to ensure privacy even against quantum computer attacks, while maximizing efficiency. To this end, we first provide an overview of stealth address techniques that have been used in previous research and present the most widely used protocol so far, the Dual-Key Stealth Address Protocol (DKSAP) \cite{textbook1}, and the most efficient protocol, ECPDKSAP (Protocol 3 from \cite{MM}).

We propose three new post-quantum protocols based on the LWE technique: LWE SAP, Ring-LWE SAP and Module-LWE SAP. The paper shows how the privacy of transaction recipients can be preserved using post-quantum cryptography. Furthermore, we compare all of our proposed protocols in terms of their efficiency, focusing on the time required for a recipient to compute the public key for their new stealth address. These comparisons also include the Curvy protocol (Protocol 3 from \cite{MM}).

The results clearly show that our Module-LWE SAP is the most efficient and outperforms the Curvy protocol. In particular, our Module-LWE SAP takes about 15 times less time to compute the recipient address compared to DKSAP and about three times less time compared to Curvy.

One observation we made while benchmarking various operations in our protocol is that the time-consuming processes are predominantly from the Key Encapsulation Mechanism (KEM) itself and not from the computation of the stealth address public key. Consequently, efforts to optimize the underlying KEM could lead to significant improvements.

Furthermore, all protocols presented in this paper are quantum resistant and are expected to play an important role in the future, especially when blockchains adopt larger key sizes and the security of current protocols becomes compromised.


\begin{thebibliography}{}

\bibitem{textbook19}Peter Todd, \emph{"[Bitcoin-development] Privacy and blockchain data"}, 2014. Available: \href{https://www.mail-archive.com/bitcoin-development@lists.sourceforge.net/msg03612.html}{https://www.mail-archive.com/bitcoin-development@lists.sourceforge.net/msg03612.html}


\bibitem{textbook20}Nicolas van Saberhagen, \emph{"CryptoNote v 2.0"}, 2013. Available: \href{https://bytecoin.org/old/whitepaper.pdf}{https://bytecoin.org/old/whitepaper.pdf}

\bibitem{ringsign}Ronald L. Rivest, Adi Shamir, Yael Tauman, \emph{"How to Leak a Secret"}, Advances in Cryptology — ASIACRYPT 2001, pp.\ 552--565, 2001, doi: \href{https://doi.org/10.1007/3-540-45682-1_32}{10.1007/3-540-45682-1\_32}

\bibitem{ringct}Shen Noether, Adam Mackenzie, \emph{"Ring Confidential Transactions"}, Ledger, 2016, doi: \href{https://doi.org/10.5195/ledger.2016.34}{10.5195/LEDGER.2016.34}

\bibitem{textbook1}Anton Wahrstatter, Matthew Solomon, Ben DiFrancesco, Vitalik Buterin, Davor Svetinovic, \emph{"BaseSAP: Modular Stealth Address Protocol for Programmable Blockchains"}, IEEE Transactions on Information Forensics and Security, vol.\ 19, pp.\ 3539--3553, 2024, doi: \href{https://ieeexplore.ieee.org/document/10426757} {10.1109/TIFS.2024.3364081}

\bibitem{textbook4}Vitalik Buterin, \emph{"An incomplete guide to stealth addresses"}, 2023. Available: \href{https://vitalik.eth.limo/general/2023/01/20/stealth.html}{https://vitalik.eth.limo/general/2023/01/20/stealth.html}

\bibitem{textbook5}Vitalik Buterin, \emph{"Possible futures of the Ethereum protocol, part 6: The Splurge"}, 2024. Available: \href{https://vitalik.eth.limo/general/2024/10/29/futures6.html}{https://vitalik.eth.limo/general/2024/10/29/futures6.html}

\bibitem{ecpsap1}Cong Feng, Liang Tan, Huan Xiao, Xin Qi, Zheng Wen, Yang Liu, \emph{"EDKSAP: Efficient Double-Key Stealth Address Protocol in Blockchain"}, IEEE 20th International Conference on Trust, Security and Privacy in Computing
and Communications (TrustCom), pp.\ 1196–1201, 2021, doi: \href{https://doi.org/10.1109/TrustCom53373.2021.00162}{10.1109/TrustCom53373.2021.00162}

\bibitem{ecpsap2}Jia Fan, Zhen Wang, Yili Luo, Jian Bai, Yarong Li, Yao Hao, \emph{"A New Stealth Address Scheme for Blockchain"}, Proceedings of the ACM Turing Celebration Conference - China, Article number 67, pp.\ 1--7, 2019, doi: \href{https://dl.acm.org/doi/10.1145/3321408.3321573}{10.1145/3321408.3321573}

\bibitem{regev}Oded Regev, \emph{"On lattices, learning with errors, random linear codes, and cryptography"}, STOC '05: Proceedings of the thirty-seventh annual ACM symposium on Theory of computing, pp.\ 84--93, 2005, doi: \href{https://doi.org/10.1145/1060590.1060603}{10.1145/1060590.1060603}

\bibitem{regev2} Vadim Lyubashevsky, 
Chris Peikert, Oded Regev, \emph{"On Ideal Lattices and Learning with Errors over Rings"}, Journal of the ACM (JACM), vol.\ 60, Issue 6, pp.\ 1--35, 2013, doi: \href{https://doi.org/10.1145/2535925}{10.1145/2535925}

\bibitem{ajtai}Miklós Ajtai, \emph{"Generating hard instances of lattice problems"}, STOC '96: Proceedings of the twenty-eighth annual ACM symposium on Theory of Computing, pp.\ 99--108, 2015, doi: \href{https://doi.org/10.1145/237814.237838}{10.1145/237814.237838}

\bibitem{lattice3}Chris Peikert, \emph{"A Decade of Lattice Cryptography"}, Foundations and Trends in Theoretical Computer Science, vol.\ 10, no.\ 4, pp.\ 283--424, 2016,  doi: \href{http://dx.doi.org/10.1561/0400000074}{10.1561/0400000074}

\bibitem{fhe-gentry} Zvika Brakerski, Craig Gentry, Vinod Vaikuntanathan, \emph{"Fully Homomorphic Encryption without Bootstrapping"}, ITCS '12: Proceedings of the 3rd Innovations in Theoretical Computer Science Conference, pp.\ 309--325, 2012, doi: \href{https://doi.org/10.1145/2090236.2090262}{10.1145/2090236.2090262}

\bibitem{lattice5}Daniele Micciancio, Shafi Goldwasser, \emph{Complexity of Lattice Problems: A Cryptographic Perspective}, Springer, 2002, doi: \href{https://link.springer.com/book/10.1007/978-1-4615-0897-7}{10.1007/978-1-4615-0897-7}

\bibitem{nist}Dustin Moody, Ray Perlner, Andrew Regenscheid, Angela Robinson, David Cooper, \emph{Transition to Post-Quantum Cryptography Standards}, NIST Internal Report (IR) NIST IR
8547 ipd, 2024, doi: \href{https://doi.org/10.6028/NIST.IR.8547.ipd}{10.6028/NIST.IR.8547.ipd} 

\bibitem{kyber}Roberto Avanzi, Joppe Bos, Léo Ducas, Eike Kiltz, Tancrède Lepoint,
Vadim Lyubashevsky, John M. Schanck, Peter Schwabe, Gregor Seiler, Damien Stehlé, \emph{CRYSTALS-Kyber
Algorithm Specifications And Supporting Documentation (version 3.01)}, January 31, 2021. Available: \href{https://pq-crystals.org/kyber/data/kyber-specification-round3-20210131.pdf}{https://pq-crystals.org/kyber/data/kyber-specification-round3-20210131.pdf}

\bibitem{dilithium}Shi Bai, Léo Ducas, Eike Kiltz, Tancrède Lepoint, Vadim Lyubashevsky, Peter Schwabe, Gregor Seiler and Damien Stehlé, \emph{CRYSTALS-Dilithium
Algorithm Specifications And Supporting Documentation (version 3.01)}, February 8, 2021. Available: \href{https://pq-crystals.org/dilithium/data/dilithium-specification-round3-20210208.pdf}{https://pq-crystals.org/dilithium/data/dilithium-specification-round3-20210208.pdf}

\bibitem{module}Martin R. Albrecht, Amit Deo, \emph{"Large Modulus Ring-LWE $\geq$ Module-LWE"}, Advances in Cryptology — ASIACRYPT 2017, pp.\ 267--296, 2017, doi: \href{http://dx.doi.org/10.1007/978-3-319-70694-8_10}{10.1007/978-3-319-70694-8\_10}

\bibitem{MM}Marija Mikić, Mihajlo Srbakoski, \emph{Elliptic Curve Pairing Stealth Address Protocols}, 2024, doi:\href{https://doi.org/10.48550/arXiv.2312.12131}{10.48550/arXiv.2312.12131} - \emph{on review}

\bibitem{MM2}\emph{"0x3327/ecpdksap"}, 2024. Available: \href{https://github.com/0x3327/ecpdksap}{https://github.com/0x3327/ecpdksap}

\bibitem{kyber-original} Joppe Bos, Léo Ducas, Eike Kiltz, Tancrède Lepoint, Vadim Lyubashevsky, John M. Schanck, Peter Schwabe, Gregor Seiler, Damien Stehlé, \emph{CRYSTALS – Kyber: a CCA-secure module-lattice-based KEM}, 2018 IEEE European Symposium on Security and Privacy (EuroS\&P), pp.\ 353--367, doi: \href{https://doi.org/10.1109/EuroSP.2018.00032}{10.1109/EuroSP.2018.00032}

\bibitem{katz-lindell}Jonathan Katz, Yehuda Lindell, \emph{Introduction to Modern Cryptography}, Chapman and Hall, 2nd edition, 2014, doi: \href{https://doi.org/10.1201/b17668}{10.1201/b17668}

\bibitem{textbook7}Eli Ben-Sasson, Alessandro Chiesa, Christina Garman, Matthew Green, Ian Miers, Eran Tromer, Madars Virza, \emph{"Zerocash: Decentralized Anonymous Payments from Bitcoin"}, IEEE Symposium on Security and Privacy, pp.\ 459--474, 2014, doi: \href{https://ieeexplore.ieee.org/document/6956581}{10.1109/SP.2014.36}

\bibitem{fhedksap}Yuping Yan, George Shao, Dennis Song, Mason Song, Yaochu Jin \emph{"HE-DKSAP: Privacy-Preserving Stealth Address Protocol via Additively Homomorphic Encryption"}, 2023, doi: \href{https://doi.org/10.48550/arXiv.2312.10698}{10.48550/arXiv.2312.10698}

\bibitem{fhe}Craig Gentry \emph{"Fully homomorphic encryption using ideal lattices"}, Proceedings of the 41st annual ACM symposium on Theory of computing, pp.\ 169--178, 2009, doi: \href{https://doi.org/10.1145/1536414.1536440}{10.1145/1536414.1536440}

\bibitem{zcash}\emph{"Zcash"}, 2016. Available: 
\href{https://z.cash}{https://z.cash}

\bibitem{tornado}\emph{"Tornado Cash"}, 2021. Available: 
\href{https://github.com/tornadocash/docs}{https://github.com/tornadocash}

\bibitem{textbook8}\emph{"Umbra Cash"}, 2021. Available: \href{https://app.umbra.cash}{https://app.umbra.cash}

\bibitem{textbook9}\emph{"Fluidkey"}, 2023. Available: \href{https://www.fluidkey.com}{https://www.fluidkey.com}

\bibitem{textbook10}\emph{"Railgun"}, 2023. Available: \href{https://railgun.org}{https://railgun.org}

\bibitem{textbook11}\emph{"Labyrinth"}, 2023. Available: \href{https://labyrinth.technology}{https://labyrinth.technology}

\bibitem{textbook12}\emph{"Monero project"}, 2014. Available: \href{https://www.getmonero.org}{https://www.getmonero.org}

\bibitem{shor} Peter W. Shor \emph{"Algorithms for quantum computation: discrete logarithms and factoring"}, Proceedings 35th Annual Symposium on Foundations of Computer Science, 1994, pp.\ 124--134, doi: \href{https://doi.org/10.1109/SFCS.1994.365700}{110.1109/SFCS.1994.365700}

\bibitem{pq-sap-repo} \emph{"0x3327/pq-sap"}, 2024. Available: \href{https://github.com/0x3327/pq-sap/tree/main/src}{https://github.com/0x3327/pq-sap/tree/main/src}

\bibitem{kyber-kem-repo} \emph{"Argyle-Software/kyber"}, 2023. Available: \href{https://github.com/Argyle-Software/kyber}{https://github.com/Argyle-Software/kyber}

\bibitem{newhope-repo} \emph{"newhopecrypto/newhope"}, 2020. Available: \href{https://github.com/newhopecrypto/newhope}{https://github.com/newhopecrypto/newhope}

\bibitem{frodo-kem-repo} \emph{"microsoft/PQCrypto-LWEKE}, 2023. Available: \href{https://github.com/microsoft/PQCrypto-LWEKE/commits/master/}{https://github.com/microsoft/PQCrypto-LWEKE/commits/master/}

\bibitem{newhope-paper} Erdem Alkim, Léo Ducas, Thomas Pöppelmann, Peter Schwabe, \emph{NewHope without reconciliation}, 2016,. Available: \href{https://eprint.iacr.org/2016/1157}{https://eprint.iacr.org/2016/1157}

\bibitem{frodo-kem-paper} Erdem Alkim, Joppe W. Bos, Léo Ducas, Patrick Longa, Ilya Mironov, Michael Naehrig, Valeria Nikolaenko, Chris Peikert, Ananth Raghunathan, Douglas Stebila, \emph{FrodoKEM Learning With Errors Key Encapsulation}, 2021. Available: \href{https://frodokem.org/files/FrodoKEM-specification-20210604.pdf}{https://frodokem.org/files/FrodoKEM-specification-20210604.pdf}
\bibitem{delta-correctness-paper} Dennis Hofheinz, Kathrin Hövelmanns, Eike Kiltz, \emph{A Modular Analysis of the Fujisaki-Okamoto Transformation}, 15th Theory of Cryptography Conference, 2017, pp.\ 341--371, doi: \href{https://doi.org/10.1007/978-3-319-70500-2_12}{10.1007/978-3-319-70500-2\_12}

\end{thebibliography}
\end{document}